\documentclass[twocolumn,aps,prl,superscriptaddress,nofootinbib,10pt]{revtex4-1}
\usepackage[latin1,utf8]{inputenc}
\usepackage{graphicx}
\usepackage{svg}
\usepackage{amsmath,amssymb,amsfonts}
\usepackage{units}
\usepackage{bm,natbib}
\usepackage{subfigure}
\usepackage{scalefnt}
\usepackage{color}
\usepackage{placeins}

\begin{document}
\title{{Effects of clays on spin-spin relaxation: a route for non-invasive total clay content quantification}}
\author{Jefferson G. Filgueiras}
\affiliation{Instituto de Química, Universidade Federal Fluminense, Outeiro de São João Batista, s/nº, 24020-007, Niterói, RJ, Brazil}
\email{jgfilgueiras@id.uff.br}
\author{Matheus S. J. de Miranda}
\affiliation{Instituto de Química, Universidade Federal Fluminense, Outeiro de São João Batista, s/nº, 24020-007, Niterói, RJ, Brazil}
\author{Carla S. Semiramis}
\affiliation{Instituto de Química, Universidade Federal Fluminense, Outeiro de São João Batista, s/nº, 24020-007, Niterói, RJ, Brazil}
\author{Rodrigo B. V. de Azeredo}
\affiliation{Instituto de Química, Universidade Federal Fluminense, Outeiro de São João Batista, s/nº, 24020-007, Niterói, RJ, Brazil}

\begin{abstract}
Clay minerals are important components of sandstone rocks, due to their significant role in petrophysical properties like porosity and permeability. These minerals have a particular impact on Nuclear Magnetic Resonance measurements since the iron inside clays generates internal gradients that impact the transverse relaxation directly. Here, we apply a methodology recently developed to a set of 20 sandstones with diverse clay content and mineralogy to estimate the total clay content. This estimation relies on the effect of internal gradients and restricted diffusion on transverse relaxation. Our analysis revealed a linear correlation between the total clay content and the displacement of the peak of the T$_2$ distribution as a function of $\tau$, which is half the echo time in the CPMG sequence. Based on these measurements, we propose a geochemical rock typing from quantities determined by our measurements, namely total clay content and porosity.
\end{abstract}


\maketitle


The identification and quantification of clay minerals play a key role in sandstone reservoir characterization, since they impact petrophysical properties like water saturation, permeability, and wettability, among others \cite{Selley2015}. Since such minerals are composed of small grains, they augment the residual water saturation due to higher capillary retention \cite{Jacomo2018,Worthington2011}. Thus, for sandstone reservoir logging, an accurate determination of the total clay content helps to improve the interpretation of data of gamma-ray and neutron porosity logs, which are particularly sensitive to the presence of clay minerals \cite{Worthington2011}.

The structure of clay minerals varies widely since they can exist in different forms and chemical compositions \cite{Moore1998}. For example, the non-swelling illite is usually found as (K,H$_3$O)(Al,Mg,Fe)$_2$(Si,Al)$_4$O$_{10}$[(OH)$_2$,(H$_2$O)]. In particular, some common clay constituents are paramagnetic ions like iron (Fe) and manganese (Mn) \cite{Brindley1980}. Despite the importance of total clay content for a precise interpretation of log data, the standard way to estimate the clay content using XRD techniques requires the samples in powder, i.e., they are invasive measurements and cannot be performed in the well.

The paramagnetic ions directly affect the magnetic susceptibility of the pore matrix. Thus, the chemical composition of the clays in a sandstone impact the Nuclear Magnetic Resonance (NMR) response \cite{Hurlimann2012,Elsayed2022a}, a technique often employed as a logging tool to evaluate oil fields. In particular, the paramagnetic ions generate internal gradients inside the pores, resulting in a relaxation mechanism observed in sandstones. As the nuclear spins diffuse inside the pores, their phases are modified by the internal gradients as they travel within the pore space, resulting in magnetization loss. This mechanism, called diffusive relaxation, occurs simultaneously with the surface relaxation resulting from the interaction between the fluid and the pore surfaces.

This diffusive relaxation reduces the transverse relaxation time, T$_2$, increasing the complexity of the NMR data interpretation. For example, the reduction in T$_2$ can be misinterpreted as the presence of smaller pores \cite{Coates1999}. However, we can estimate the internal gradients by different techniques, since this relaxation mechanism depends explicitly on the echo time used in the T$_2$ measurements. 1D and 2D pulse sequences have been proposed in the literature, along with the discussion of how to interpret such data due to assumptions made during data analysis \cite{Hurlimann1998,Hurlimann2002,Sun2002,Appel1999,Dunn2001}. Recently, Elsayed and collaborators proposed a simple way to estimate the total clay content using the diffusive relaxation mechanism  \cite{Elsayed2020}. On their method, only two T$_2$ measurements are necessary to measure the clay content, which can be done \emph{in situ} during drilling. Using only seven samples, they observed the connection between total clay content and the relative reduction of T$_2$ due to diffusive relaxation, showing a nonlinear relation between these two variables.

In this paper, we test the hypothesis by Elsayed \emph{et al} in a set of 20 sandstones, with varying total clay content and diverse mineral compositions. Instead of a nonlinear behavior, we observe a linear correlation between total clay content and the relative reduction of T$_2$. We also discuss how some specific minerals, like ankerite and kaolinite, can impact the total clay estimation using diffusive relaxation, and we propose a simple rock typing based on the total clay content.  


\section{Theory} 

Time-domain NMR measures the magnetization dynamics of $^1$H nuclear spins when removed from equilibrium. Since the state of the nuclear spins is directly affected by the molecular mobility, the NMR response is sensitive to the environment surrounding the nuclear spins of the liquid used as molecular probe \cite{Callaghan2014,Abragam1961}. In particular, the transverse relaxation time, T$_2$, can be used to study the geometry of porous media, since it is affected by the interaction of the saturating fluid with the pore walls. This relaxation mechanism is the surface relaxation and, in the fast diffusion limit, directly relates T$_2$ and pore size \cite{Callaghan2014}. Since the magnetization is directly proportional to the amount of fluid, the T$_2$ measurement also allows us to estimate the pore volume (provided the sample is fully saturated), i.e., the effective porosity. For this reason, NMR measurements are widely adopted to estimate properties like porosity and permeability in reservoir rocks \cite{Hurlimann2012}. In parallel to the surface relaxation mechanism, when there are internal gradients in a rock, the nuclear spins diffusing on such gradients accumulate a phase dependent on both the gradient intensities and the positions of the molecules. Since this phase cannot be fully refocused after a 180$^o$ spin inversion pulse in an echo sequence, there is a decrease in T$_2$ due to this diffusion effect. Thus, the transverse relaxation T$_2$ can be described by \cite{Callaghan2014}

\begin{equation}\label{T2}
\frac{1}{T_2} = \frac{1}{T_{2}^{bulk}} + \rho_2\frac{S}{V} + \frac{1}{3}\gamma^2G^2D_0\tau^2,
\end{equation}
\par\noindent where T$_{2}^{bulk}$ is the bulk relaxation time of the fluid, $\rho_2$ is the surface relaxivity, $S$ and $V$ are the pore surface and volume, respectively. The last term in the expression describes the effect of the diffusive relaxation, with $\gamma$ being the gyromagnetic ratio for the proton, $G$ being the internal field, and $D_0$ the self-diffusion coefficient of the fluid. Finally, $\tau$ is half the echo time in the Carr-Purcell-Meiboom-Gill (CPMG) sequence \cite{Carr1954,Meiboom1958}. This equation is valid if we assume a fast diffusion regime for surface relaxation and free diffusion for diffusive relaxation \cite{Callaghan2014, Dunn2001}. When we consider porous media, the last assumption can lead to some problems in interpreting the data, since it is valid only in the case of a uniform fluid, i.e., without any effects of restricted diffusion. However, it was demonstrated that such an assumption is reasonable to describe the internal fields in sandstone rocks \cite{Dunn2001}.

Equation \eqref{T2} describes how the relaxation time decreases due to the effect of internal gradients when we increase the echo time. Therefore, we can gain information on the internal gradients by measuring the T$_2$ distribution using several $\tau$ values. Using such analysis it is possible to recover the distribution of effective internal gradients of a porous media \cite{Hurlimann1998} and even correlate the effective internal gradient to pore size using a 2D experiment \cite{Hurlimann2002,Sun2002}. Here, we use Eq. \eqref{T2} to observe how the relaxation rate (${1}/{T_2}$) behaves as a function of $\tau^2$. For short diffusion times, the spins do not interact with pore surfaces, the free diffusion is valid and we observe a linear behavior. However, as $\tau^2$ increases, there is a deviation from the linear behavior due to restricted diffusion \cite{Appel1999,Dunn2001,Elsayed2020}, reflected in a reduction of the apparent diffusion coefficient. This restriction effect is notably strong on small pores like the ones associated with clays. 

For sandstones, the internal gradient field is related mainly to paramagnetic ions on the pore matrix, usually iron. So, we should expect an increase in the intensity of the internal gradients with higher clay content. As such, Elsayed et al hypothesized that any clay in a porous medium correlates to shifts in T$_2$ \cite{Elsayed2020}. 

\section{Materials and methods} 

\textbf{Rock samples.} We use 20 sandstone rocks with a 1.5" diameter and 5 cm length, provided by Kocureck Industries Caldwell, USA). We cleaned the samples with Soxhlet extraction to remove any hydrocarbons or salts inside the pores. We dried the plugs at 60$^o$C in an oven. We measured both porosity and permeability using a porosimeter and a permeater. We calculated porosity using Boyle's law and permeability using Darcy's law. The list of the samples, including their porosity and permeability, is shown in Table \ref{tab:1}. In particular, the Idaho Gray sample has a permeability higher than the sensibility limit of the permeameter (5000 mD).

\begin{table*}[t]
    \centering
    \caption{\label{tab:1} \textbf{Petrophysical and mineralogical characterization.} Porosity and permeability of the 20 sandstone samples. We determined their respective mineralogical composition by using XRD with the Rietveld method. We measured the Fe \%wt with X-ray fluorescence. The error in the amount of each mineral is 1.0 \%wt. For the total clay content the error is 1.7 \%wt. ($\phi$: porosity; $\phi_{NMR}$: NMR porosity; $K$ (mD): absolute permeability; qtz: quartz (SiO$_2$); feld: sum of K-feldspars and plagioclause; ill: sum of illite, biotite and muscovite; chlo: chlorite; kaol: kaolinite; oxi: sum of iron oxides (magnetite and hematite); zeo: zeolite (clinoptilolite); dol: dolomite; ank: ankerite; cal: calcite; Fe: weight \% of Fe on each sample; $\sum$clays: total clay content)}
    \resizebox{\textwidth}{!}{
    \begin{tabular}{c c c c c c c c c c c c c c c c}
        \hline
         & $\phi$ & $\phi_{NMR}$ & $K$ (mD) & qtz & feld & ill/mica & chlo & kaol & oxi & zeo & dol & ank & cal & \% Fe &  $\sum$clays \\ \hline 
        
        Bandera Brown & 20.7 & 20.4 & 1.0 & 67.3 & 16.0 & 8.1 & 2.8 & 4.1 & nd &nd & nd & nd & 1.7 & 7.0 & 15.0\\
        
        Berea Stripe & 20.0 & 19.7 & 368 & 88.5 & 4.0 & 2.2 & nd & 2.8 & nd & nd & nd & 2.6 & nd & 2.7 & 5.0 \\ 
        
        Kirby & 21.2 & 20.6 & 13 & 79.7 & 12.7 & 3.3 & nd & 4.2 & nd & nd & nd & nd & nd & 2.2 & 7.5\\ 
        
        Bentheimer & 22.6 & 22.5 & 2805 & 94.1 & 3.8 & nd & nd & 2.1 & nd & nd & nd & nd & nd & 0.2 & 2.1  \\

        Leopard & 21.1 & 19.0 & 1683 & 96.3 & nd & 2.1 & nd & 1.5 & 0.1 & nd & nd & nd & nd & 1.7 & 3.6  \\

        Briarhill & 25.5 & 23.3 & 4842 & 92.9 & 3.5 & 0.6 & nd & 2.6 & nd & nd & nd & nd & nd & 0.8 & 3.2 \\

        Torey Buff & 17.0 & 16.9 & 1.4 & 54.4 & 8.1 & 3.0 & nd & 10.1 & nd & nd & 9.6 & 14.8 & nd & 7.7 & 13.1  \\

        Boise Idaho Brown & 27.9 & 27.3 & 1739 & 38.9 & 49.9 & 3.1 & nd & nd & nd & 8.1 & nd & nd & nd & 2.7 & 3.1  \\

        Buff Berea & 24.5 & 23.5 & 668 & 85.2 & 9.1 & 1.5 & nd & 4.2 & nd & nd &  nd & nd & nd & 3.8 & 5.7  \\

        Berea & 19.9 & 19.5 & 149 & 85.7 & 7.7 & 2.3 & 1.2 & 1.9 & nd & nd & nd & 1.2 & nd & 4.0 & 5.4 \\

        Bandera Gray & 21.1 & 20.7 & 17.7 & 61.6 & 14.2 & 6.4 & 5.3 & 2.3 & 0.8 & nd & nd & 8.1 & 1.3 & 7.6 & 14.0  \\

        Castlegate & 26.0 & 24.3 & 1147 & 90.0 & 5.8 & 2.0 & 0.4 & 2.0 & nd & nd & nd & nd & nd & 1.8 & 4.4  \\

        Idaho Gray & 27.9 & 27.6 & $>$5000 & 46.0 & 48.7 & 1.8 & nd & nd & nd & 3.3 & nd & nd & nd & 1.3 & 1.8 \\

        Boise Idaho Gray & 29.6 & 29.1 & 4628 & 45.0 & 48.1 & 3.6 & nd & nd & nd & 3.3 & nd & nd & nd & 1.5 & 3.6  \\

        Kentucky & 14.1 & 14.9 & 0.4 & 63.7 & 23.9 & 9.8 & 2.2 & nd & 0.3 & nd & nd & nd & nd & 5.8 & 12.0  \\

        Nugget & 10.8 & 10.0 & 8.8 & 82.9 & 10.3 & 3.7 & nd & 2.2 & 0.4 & nd & 0.5 & nd & nd & 2.1 & 5.9 \\

        Scioto & 15.8 & 16.2 & 0.5 & 71.5 & 14.6 & 8.9 & 2.3 & 2.1 & nd & nd & 0.5 & nd & nd & 6.3 & 13.3  \\

        Sister Gray Berea & 20.4 & 19.4 & 105 & 82.8 & 10.3 & 2.8 & 1.1 & 2.2 & nd & nd & 0.8 & nd & nd & 2.8 & 6.1  \\

        Gray Berea & 19.7 & 19.6 & 201 & 80.8 & 9.8 & 3.1 & nd & 4.4 & nd & nd & nd & 1.2 & 0.6 & 3.7 & 7.5 \\

        Upper Gray Berea  & 19.1 & 18.7 & 126 & 87.0 & 6.3 & 3.0 & 0.5 & 2.4 & nd & nd & nd & 0.9 & nd & 4.0 & 5.9  \\
        \hline
        \end{tabular}
    }
\end{table*}

\textbf{Geochemistry and mineralogy.} We determined the total geochemical composition using X-ray fluorescence, with an EDXRF (PAnalytical, Malvern, UK). We assessed the mineralogical composition applying X-ray diffraction, with a D2-PHASER (Bruker, Karlsruhe, Germany). We used the following parameters in the measurements: range  3$^o$ to 100$^o$, 0.02$^o$ step size, and a 3 s scanning time. The XRD data was processed using the EVA\textsuperscript{\tiny\textregistered} software. For the mineralogical quantification of the samples, we followed the Rietveld Method using DIFFRAC.SUITE TOPAS\textsuperscript{\tiny\textregistered} software \cite{Mccusker1999,Toby2006}. We grouped plagioclase and alkali feldspar together, and summed the mica structures (illite, biotite, and muscovite). Ankerite is one type of dolomite structure, with low or high Fe content. The uncertainty in each mineral is 1 \%wt, and the error for total clay content is 1.7 \%wt, obtained through error propagation.

\textbf{Nuclear Magnetic Resonance.} We performed $^1$H NMR relaxometry measurements using a Geospec DRX core analyzer (Oxford Instruments, Oxford, UK), with a Larmor frequency of 2.2 MHz (0.05 T), with 11.7 and 23.6 $\mu$s for the 90$^o$ and 180$^o$ pulse widths, respectively. We saturated the samples with a KCl brine at 30000 ppm concentration and acquired CPMG measurements with 21 geometrically spaced echo times between 0.2 and 6 milliseconds using the GIT software (Green Image Technologies, Fredericton, Canada). We obtained the T$_2$ distributions using an Inverse Laplace Transform (ILT) for each echo time, with 512 points. We wrapped the core samples with 2" wide Teflon tape and put them inside a PEEK support to prevent fluid loss during the measurements. The NMR porosity was estimated using a standard sample with known water volume, provided by Green Image Technologies.

\section{Results} 

\begin{table*}[t!]
    \centering
    \caption{\label{tab:2} \textbf{NMR clay content quantification.} The first two columns are the total clay content
    determined by XRD and the T$_{2}^{cutoff}$ method by NMR \cite{Coates1999}. The other columns indicate the values of 
    $\Delta T_{2}^{\tau}$ for the three $\tau$ values we used (1.5, 1.8, 2.2 and 2.6 ms, respectively). The last column is 
    the internal gradient field measured by using Eq. (\eqref{T2}).}
    \resizebox{1.8\columnwidth}{!}{
    \begin{tabular}{c c c c c c c c}
        \hline
         & $\sum$clays & clay bound water & $\Delta T_{2}^{1.5 ms}$ & $\Delta T_{2}^{1.8 ms}$ & $\Delta T_{2}^{2.2 ms}$ & $\Delta T_{2}^{2.6 ms}$ &  G (G/cm)\\ \hline 
        
        Bandera Brown & 15.0 & 13.8 & 0.81. & 0.77 & 0.77 & 0.72 & 109(12) \\
        
        Berea Stripe & 5.0 & 1.9 & 0.17 & 0.17 & 0.24 & 0.24 & 17.8(2.2) \\ 
        
        Kirby & 7.5 & 6.6 & 0.28 & 0.30 & 0.37 & 0.40 & 31.5(6.7) \\ 
        
        Bentheimer & 2.1 & 0.8 & 0.15 & 0.17 & 0.17 & 0.17 & 6.7(0.7) \\

        Leopard & 3.6 & 10.2 & 0.13 & 0.15 & 0.15 & 0.19 & 11.4(2.9) \\

        Briarhill & 3.2 & 0.3 & 0.20 & 0.20 & 0.24 & 0.22 & 7.0(0.8) \\

        Torey Buff & 13.1 & 11.2 & 0.40 & 0.39 & 0.37 & 0.28 & 113(17) \\

        Boise Idaho Brown & 3.1 & 6.7 & 0.26 & 0.28 & 0.32 & 0.35 & 17.1(1.4) \\

        Buff Berea & 5.7 & 1.9 & 0.13 & 0.15 & 0.17 & 0.20 & 19.4(1.9) \\

        Berea & 5.4 & 2.4 & 0.33 & 0.37 & 0.43 & 0.40 & 37.2(3.0) \\

        Bandera Gray & 14.0 & 7.6 & 0.61. & 0.69 & 0.73 & 0.80 & 83.5(9.3) \\

        Castlegate & 4.4 & 2.4 & 0.10 & 0.10 & 0.15 & 0.14 & 10.0(1.2) \\

        Idaho Gray & 1.8 & 2.7 & 0.17 & 0.20 & 0.24 & 0.20 & 13.6(1.8) \\

        Boise Idaho Gray & 3.6 & 2.7 & 0.16 & 0.16 & 0.20 & 0.16 & 9.5(1.5) \\

        Kentucky & 12.0 & 14.9 & 0.68 & 0.65 & 0.62 & 0.60 & 143.4(4.3) \\

        Nugget & 5.9 & 14.4 & 0.22 & 0.32 & 0.30 & 0.35 & 31.0(7.8) \\

        Scioto & 13.3 & 9.5 & 0.68 & 0.68 & 0.64 & 0.60 & 133.8(6.6) \\

        Sister Gray Berea & 6.1 & 3.0 & 0.26 & 0.30 & 0.35 & 0.40 & 27.9(3.2) \\

        Gray Berea & 7.5 & 2.3 & 0.35 & 0.37 & 0.42 & 0.46 & 27.0(4.0) \\

        Upper Gray Berea  & 5.9 & 3.3 & 0.32 & 0.37 & 0.42 & 0.39 & 27.8(3.2) \\
        \hline
        \end{tabular}
}
\end{table*}

\textbf{Rock sample geochemistry and mineralogy.} The sandstones were classified in three groups: low ($<$ 5 \%), medium (between 5 and 10 \%) and high ($>$ 10 \%) clay content. The low clay content sandstones are Boise Idaho Gray, Idaho Gray, Bentheimer, Idaho Brown, Briarhill, Leopard, and Castlegate. The medium clay content sandstones are Berea Stripe, Berea, Buff Berea, Nugget, Upper Gray Berea, Sister Gray Berea, Gray Berea, and Kirby. The high clay-content sandstones are Kentucky, Torey Buff, Scioto, Bandera Gray, and Bandera Brown. There are two main types of crystalline cementation identified in the samples: clay (kaolinite, illite, and chlorite) and carbonatic (calcite, dolomite, and ankerite) since we could not separate quartz/feldspar overgrowth from grains. The cement varies between 2 and 24 \%, occurring mainly as clays, except the Torey Buff sample, which has intense carbonatic cementation of more than 20 \% in weight. The iron bearing minerals are the clays illite and chlorite, the carbonate ankerite, and the iron oxides magnetite and magnetite. These are the minerals responsible for the internal gradients observed in the samples. However, the oxides are present as traces, , and probably scattered throughout the rock, only in the samples Leopard, Gray Berea, Bandera Gray, and Nugget. Kaolinite is the only clay with no iron in its composition, and, as such, it does not generate any internal gradients. 

Iron is the most common paramagnetic ion found in sandstones and the main source of internal gradients in this type of rock. Table \ref{tab:1} shows each samples iron amount. While the amount of iron is important for the intensity of the internal gradients, the key feature is where this iron is placed in the pores. An iron ion far from any pore surface does not contribute to the internal field gradients, since the hyperfine interaction between the electronic spin of such ions with the nuclear spins of the saturating fluid occurs only in a very short range. Thus, iron in minerals such as ankerite, a carbonatic cement found in several of our samples, does not contribute to the internal gradients as much as iron contained in chlorite, a clay usually observed coating the pore surface. The amount of iron and the mineralogical composition of the 20 sandstones we analyzed are found in Table \ref{tab:1}. In particular, the Boise Idaho Brown sample has a large zeolite concentration. While pores in both zeolites and clays have pore sizes on the same length scale and several similarities in their structure \cite{Bish2013}, zeolites do not impact the permeability inside the pore matrix like clay minerals.

 \begin{figure}
	\centering
	\includegraphics[width=0.95\columnwidth]{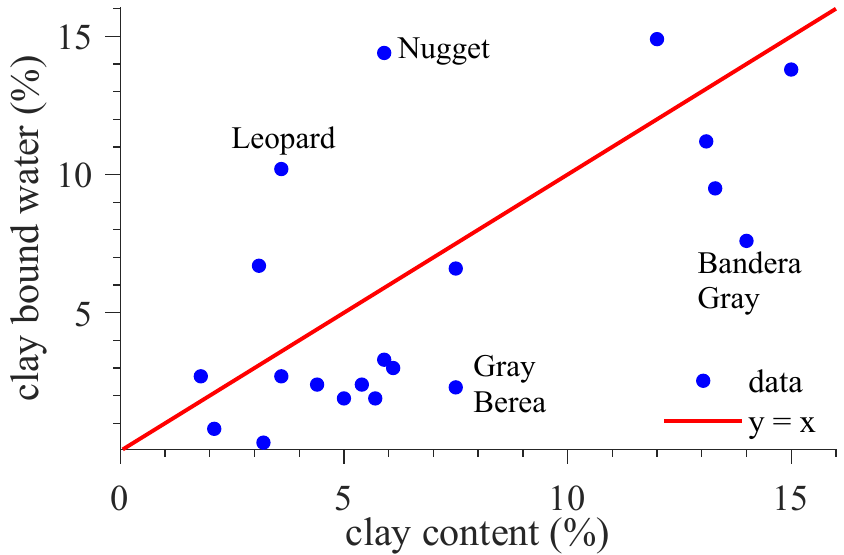}
	\caption{(Color online). Comparison by total clay content estimated using the XRD Rietveld method and using a T$_{2}^{cutoff}$ 
    of 3 ms. The clay content can be overestimated for samples with strong internal gradients or with large surface relaxivities.}
	\label{FIG:cutoff}
\end{figure}


\begin{figure*}
	\centering
	\includegraphics[width=0.8\textwidth]{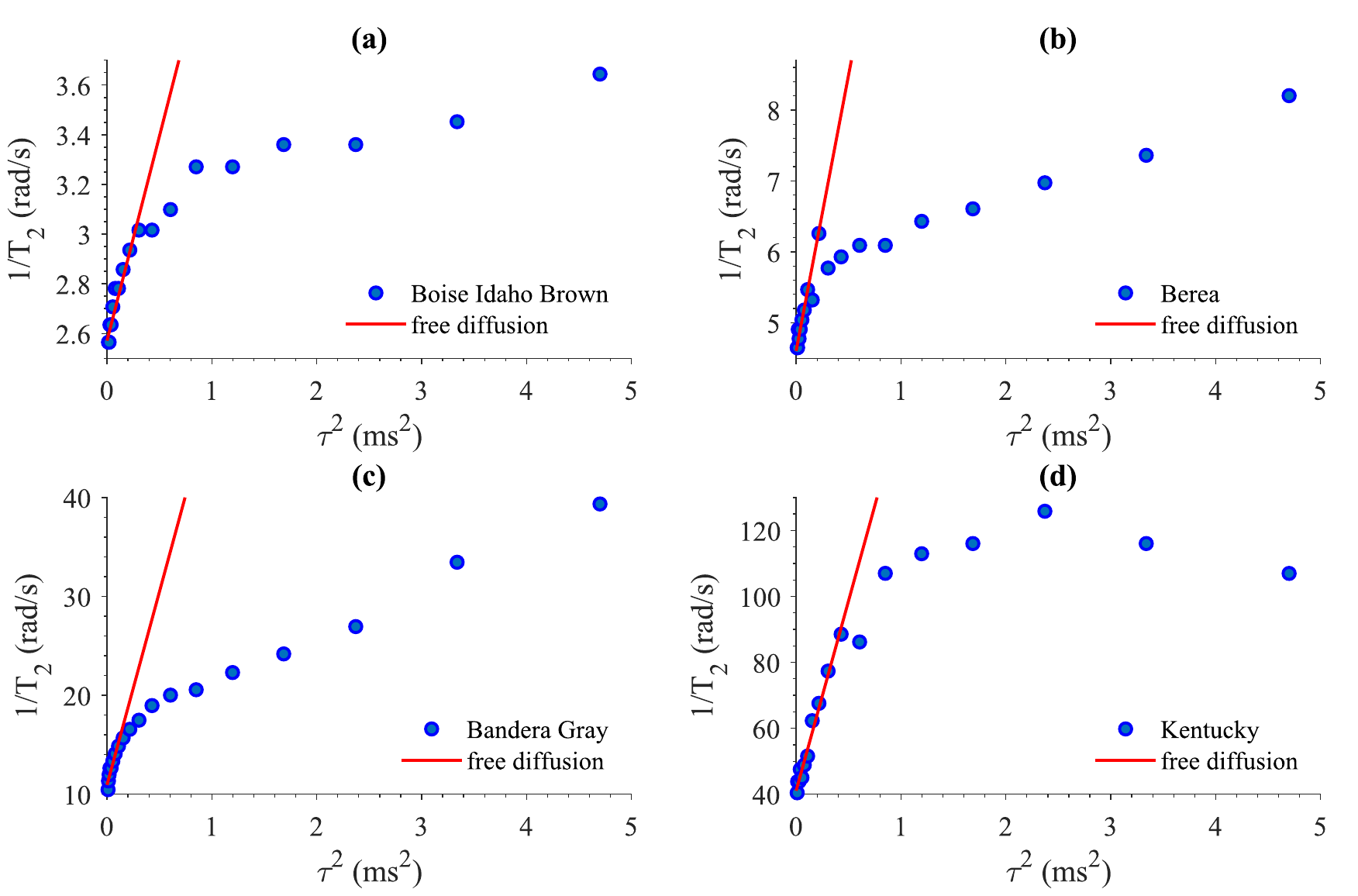}
	\caption{(Color online). Displacement of the peak of the T$_2$ distribution as a function of $\tau^2$ for the 
 samples Boise Idaho Brown (a), Berea (b),  Bandera Gray (c), and Kentucky (d). The solid line indicates the linear 
 fit for the free diffusion regime, observed for short $\tau$ values. As $\tau$ increases, we see a deviation from 
 this linear behavior due to restricted diffusion.}
	\label{FIG:fits}
\end{figure*}

\textbf{Diffusive relaxation and total clay content quantification.} The T$_2$ distribution offers a lot of information about the porous medium, such as the effective porosity and how the fluid is distributed within the pore network. In particular, it can estimate the clay content setting a T$_{2}^{cutoff}$ for clay-bound water, and the cumulative porosity below this cutoff is assumed as the clay-bound water saturation for sandstones \cite{Pramer1996,Coates1999}. While this simple measurement is readily accessible on-site, the shortest T$_2$ values are the most affected by the presence of internal gradients, and their effect can mislead the estimation of the clay content. Using this method with a T$_{2}^{cutoff}$ $=$ $3$ ms, we found a poor correlation between the total clay content determined by XRD and the clay-bound water determined by NMR. We show this data in Fig. \ref{FIG:cutoff} and Table \ref{tab:2}. We exhibit all T$_2$ distributions in the Appendix.

The low correlation observed, quantified by the Pearson correlation coefficient R$^2$ = 0.66, can be understood by looking at some of the outliers indicated in Figure 1. The samples Leopard and Nugget have 3.6 and 5.9 \% total clay content, respectively. However, their respective clay-bound water saturations of 10.2 and 14.4 \% are very large. This is due to the large surface relaxivities of these samples \cite{LucasOliveira2020}, resulting in porosity observed in the range below the T$_{2}^{cutoff}$ of 3 ms that is not associated with clays. Most samples have their clay content underestimated, especially the Bandera Gray and Gray Berea samples. For the latter, kaolinite is the dominant clay type, which has a large specific area compared to other clays, leading to a longer T$_2$ when contrasted to other clays \cite{Pramer1996}.

As already mentioned, an alternative to the T$_{2}^{cutoff}$ method has been proposed recently by Elsayed \emph{et al} \cite{Elsayed2020}. In this approach, we use the shift on the peak of the T$_2$ distribution due to the internal field gradients, which follows Eq. \eqref{T2} for short diffusion times, when the free diffusion approximation is valid. The core of the method is that the effects of restriction make the displacement of T$_{2}^{peak}$ smaller than it should be in the case of free diffusion. Since clays are usually minerals with high magnetic susceptibilities, a large amount of clays implies bigger shifts of T$_{2}^{peak}$ due to stronger internal gradients. As a figure of merit, we use the relative displacement of T$_{2}^{peak}$ in relation to the shortest $\tau$ available, of 100 $\mu$s in our case. We quantify it with the equation

\begin{equation}\label{deltaT2}
\Delta T_{2}^{\tau} = \frac{T_{2}^{0.1 ms} - T_{2}^{\tau}}{T_{2}^{0.1 ms}},
\end{equation}

\par\noindent where $T_{2}^{0.1 ms}$ is the peak of the T$_2$ distribution for $\tau$ $=$ 100 $\mu$s and $T_{2}^{\tau}$ is the peak of the distribution when half echo time is $\tau$. We emphasize that, for clay quantification, we must use $\tau$ long enough to be outside the free diffusion limit \cite{Elsayed2020}. To avoid variations on T$_{2}^{peak}$ due to redundancies in the T$_2$ distributions, we must use a large number of bins in the T$_2$ distribution. In our case, we used 512 points on each T$_2$ distribution. We used 21 geometrically spaced echo times for each of our 20 samples. We used the behavior for small $\tau^2$, i.e., the free diffusion regime, to estimate the internal gradient for each sandstone using a linear fit in function of $\tau^2$, as shown in Figure \ref{FIG:fits} for Boise Idaho Brown, Berea, Bandera Gray, and Kentucky. The fitted values for the internal gradients are shown in Table \ref{tab:2}, exhibiting a good correlation with the total iron content of the samples, as shown in the Appendix. For higher clay contents, we observe significant variations in $1/T_2$ and strong internal gradients, reflected by the slope for the free diffusion regime.

We used four $\tau$ values in order to study the behavior of $\Delta T_{2}^{\tau}$ as a linear function of the total clay content, $\tau$ $=$ 1.5, 1.8, 2.2 and 2.6 ms. We observe the optimal value for $\tau$ is 1.5 ms, with R$^{2}$ $=$ 0.90 between $\Delta T_{2}^{\tau}$ and total clay content data sets. It has a slightly better performance than $\tau$ $=$ 1.8 ms (R$^2$ $=$ 0.89), and the correlation decreases for longer $\tau$. Despite this decrease, we still observe a much better correlation when compared with the T$_{2}^{cutoff}$ method. Figure 3a shows a reasonable linear correlation between $\Delta T_{2}^{\tau}$ and the total clay content. We remark that if we use a logistic function to describe the relation between $\Delta T_{2}^{\tau}$ and the total clay content, we have almost the same correlation coefficient, R$^2$ $=$ 0.90. Thus, for simplicity, we used a linear model. One argument favorable to the logistic function is that it is a bounded function, just like $\Delta T_{2}^{\tau}$. However, since it is unusual to observe sandstones with more than 20 \% of total clay content, a linear model is sufficient to describe this relationship. 

One particularly interesting sample is Torey Buff, an outlier in Fig. 3a below the fitted line. This happens because kaolinite is the most abundant clay in its mineralogical composition, with 10.1 wt \%. As already mentioned, kaolinite is the only clay observed in our samples without any paramagnetic ion and thus does not contribute to the internal gradients. Such a feature results in a smaller shift of T$_2$ due to the internal gradients, which implies a smaller clay content predicted by our measure, $\Delta T_{2}^{\tau}$. However, as it is clear from Fig. 3a, the usage of internal fields to quantify the total clay content offers a significant advantage over the estimation of clay-bound water based on T$_{2}^{cutoff}$ shown in Fig. \ref{FIG:cutoff}.

\begin{figure}
	\centering
	\includegraphics[width=0.95\columnwidth]{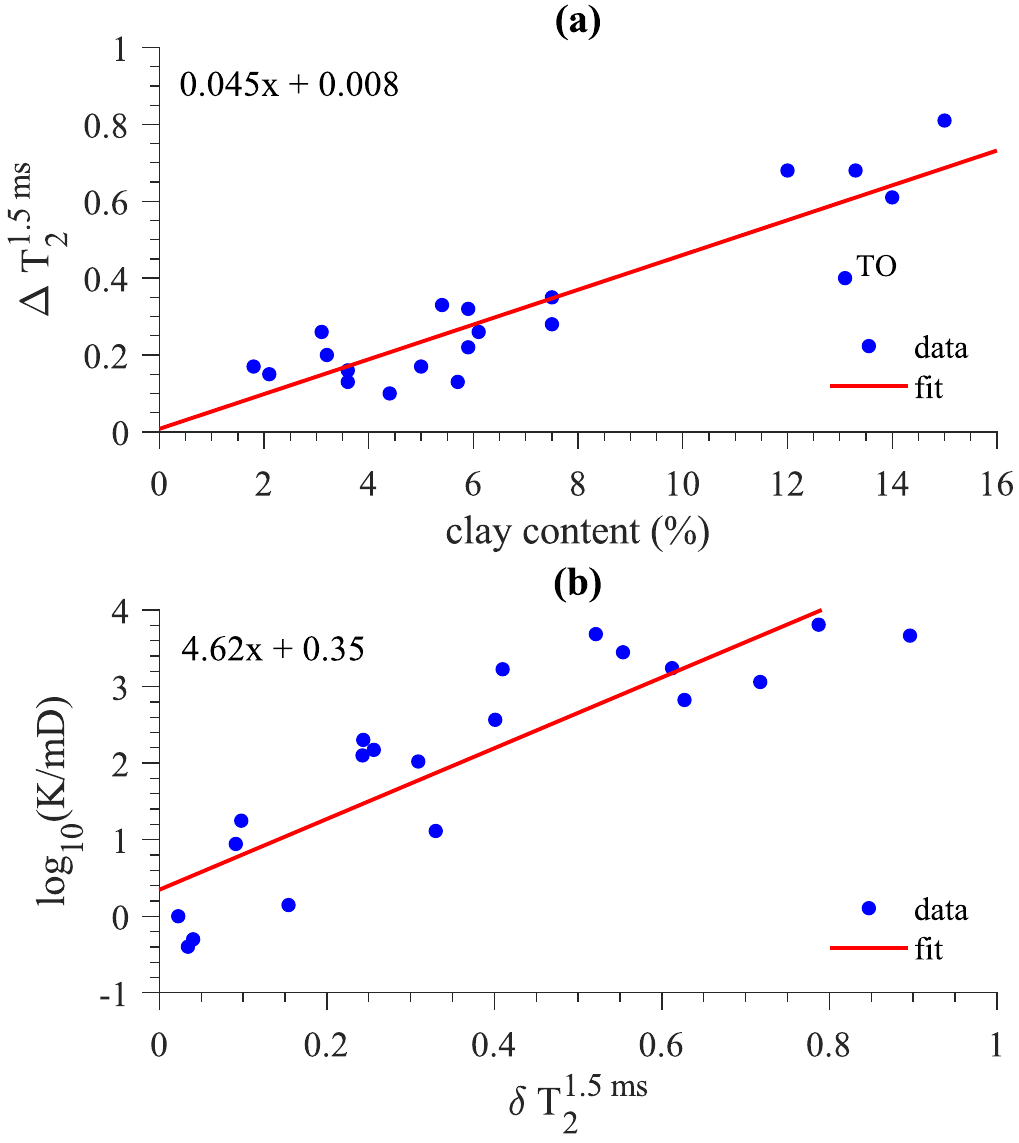}
	\caption{(Color online). Clay quantification using internal gradients. (a) $\Delta T_{2}^{\tau}$ as a linear function 
    of the total clay content. (b) Correlation between the logarithm of the permeability and $\delta$, defined in Eq. \eqref{delta}, that can be 
used as a rock typing based on the effect of clays on the spin-spin relaxation.}
	\label{FIG:quant}
\end{figure}

We can also use the total clay content to propose a simple rock typing, which could, in principle, indicate the order of magnitude of the permeability. This relies solely on the fact that clays are one of the biggest hindrances to fluid flow throughout the pore network \cite{Coates1999}. To do that, we use a slight modification on $\Delta T_{2}^{\tau}$:

\begin{equation}\label{delta}
\delta  = a\phi^2\left(1 - \Delta T_{2}^{\tau}\right)^2,
\end{equation}
\par\noindent in such a way that we have the logarithm of the permeability $K$ increasing with $\delta$ and scaled with the porosity $\phi$ and $a$ is simply a scaling factor to make $\delta$ vary between 0 and 1. We used the squares of $(1 - \Delta T_{2}^{\tau})$ to maximize the correlation between $\delta$ and log$_{10}$(K). Figure 3b shows how log$_{10}$(K) correlates well with $\delta$ (R$^2$ = 0.89), with large values for $\delta$ associated with low total clay content and the opposite holding for small $\delta$ values. By doing so, with the same measurement we can also roughly infer the sample permeability without relying on the SDR or Timur-Coates permeability models, which require the calibration of lithological constants \cite{Hurlimann2012, Elsayed2022}. In simple words, a small $\delta$ value, i.e., a large internal gradient, implies low permeability due to the high clay content. The inverse holds for $\delta\approx 1$ relating to samples with high permeability and small internal gradients.

We can use Eq. \eqref{delta} to classify the rocks into three different sets: low, medium, and high total clay content. As we see in Figure 3b, for $\delta < 0.2$, we have samples with high total clay content, bigger than 10 \%. We classify such samples as having high clay content. These samples have their permeability distributed along the two decades with the lowest permeability values shown in Figure 3b. We define as low clay content the samples with $\delta > 0.5$. Such samples have their permeability values all above 400 mD and their clay content ranging between 0 and 6 \%. For $0.2\leq\delta\le 0.5$, we define the medium total clay content rocks. In this case, the permeability varies over almost three decades, between 10 and 400 mD, and the total clay content ranges between 5 and 9 \%. We remark how $\delta$ separates the samples in three disjoint sets, as observed in Fig. 3b. This feature does not happen when we use $\Delta T_{2}^{\tau}$, as the sets with low and medium total clay content occupy the same region of $\Delta T_{2}^{\tau}$ values.

\begin{figure}
	\centering
	\includegraphics[width=0.95\columnwidth]{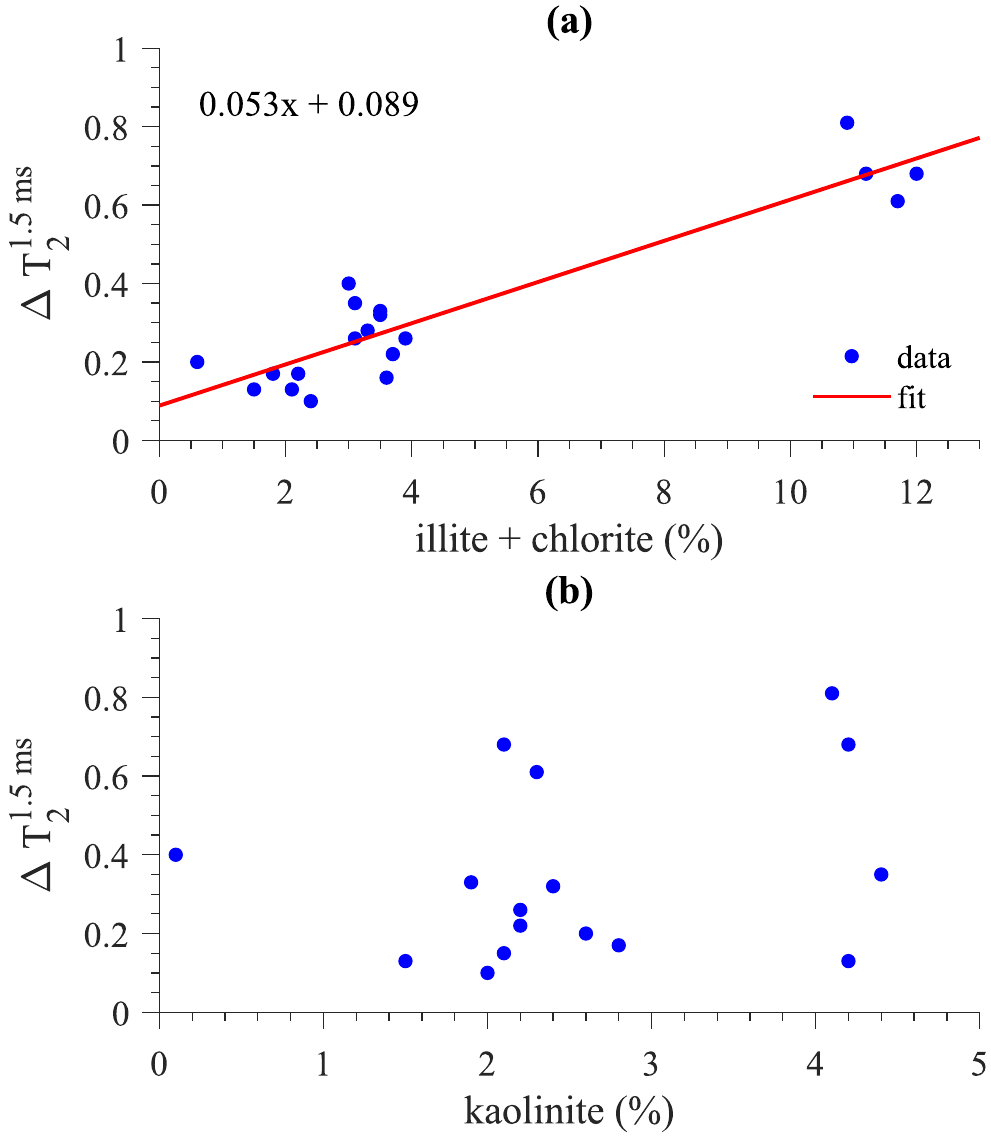}
	\caption{(Color online). Effect of clay type on $\Delta T_{2}^{\tau}$. (a) $\Delta T_{2}^{\tau}$ as a function 
    of the sum of illite and chlorite. (b) $\Delta T_{2}^{\tau}$ as a function of the kaolinite content.}
	\label{FIG:type}
\end{figure}

Finally, we evaluate how kaolinite and the sum of illite and chlorite affect the behavior of $\Delta T_{2}^{\tau}$, as shown in Fig. 4. For illite and chlorite, the clays containing iron on their structures, we observe similar behavior for $\Delta T_{2}^{\tau}$ as a function of the sum of these clays. This behavior is expected, since these two types are the ones that generate internal gradients and contribute to the diffusive relaxation mechanism. We also observe two well-separated groups for low and high content of illite and chlorite on $\Delta T_{2}^{\tau}$. For kaolinite, there is no direct relation between its content and $\Delta T_{2}^{\tau}$, a consequence of this clay not generating any internal gradients.

\section{Discussion and conclusion}

Clay quantification is an important task during well logging since an accurate estimate of the total clay content is crucial to interpreting gamma-ray and neutron porosity logs \cite{Worthington2011}. Moreover, clay cement reduces relevant properties such as porosity and permeability. Here, we applied the methodology recently developed by Elsayed and collaborators \cite{Elsayed2020} to a set of 20 sandstones, a higher number than the one used in their original work. Using such a set of samples, we observed a linear correlation between the relative displacement of the dominant peak of the T$_2$ distribution, given by Eq. \eqref{deltaT2}, and the total clay content. While such a model is distinct from the nonlinear behavior observed by Elsayed \emph{et al.}, a linear model is much simpler and easier to interpret. 

We took one step further and proposed a geochemical rock typing based on $\Delta T_{2}^{\tau}$, which allows us to determine if a rock has low, medium, or high clay content based on NMR measurements, readily available during well logging and with minimal data processing. Finally, we also showed that if we use porosity, we can define a quantity that classifies sandstones according to their permeability, as shown in Figure 3b. This simple figure of merit indicates the possibility of including the total clay content on NMR permeability models, by modifying the well-known and widely applied Timur-Coates and SDR models.

The high correlation between $\Delta T_{2}^{\tau}$ and the total clay content observed here indicates a non-invasive route for clay quantification. For \emph{in situ} applications, future work points to evaluate this methodology when dealing with more than one fluid and under different wettability conditions, since the internal gradients are particularly strong close to the pore surface.



\section*{Acknowledgements}
This research was carried out in association with the ongoing R\&D project registered as ANP nº 21289-4, "Desenvolvimento de modelos matemáticos, estatísticos e computacionais para o aperfeiçoamento da caracterização petrofísica de reservatórios por Ressonância Nuclear Magnética (RMN)” (UFF/Shell Brasil/ANP), sponsored by Shell Brasil Petróleo Ltda under the ANP R\&D levy as "Compromisso de Investimentos com Pesquisa e Desenvolvimento." The authors also recognize the support from CAPES, CNPq and FAPERJ.


\setcounter{equation}{0}
\setcounter{table}{0}
\setcounter{section}{0}
\setcounter{figure}{0}
\numberwithin{equation}{section}
\makeatletter
\renewcommand{\thesection}{\Alph{section}} 
\renewcommand{\thesubsection}{\Alph{section}.\arabic{subsection}}
\def\@gobbleappendixname#1\csname thesubsection\endcsname{\Alph{section}.\arabic{subsection}}
\renewcommand{\theequation}{\Alph{section}\arabic{equation}}
\renewcommand{\thefigure}{\arabic{figure}}
\renewcommand{\bibnumfmt}[1]{[#1]}
\renewcommand{\citenumfont}[1]{#1}
\renewcommand{\thefigure}{A\arabic{figure}}

\section*{Appendix}


\begin{figure}[b!]
	\centering
	\includegraphics[width=0.9\columnwidth]{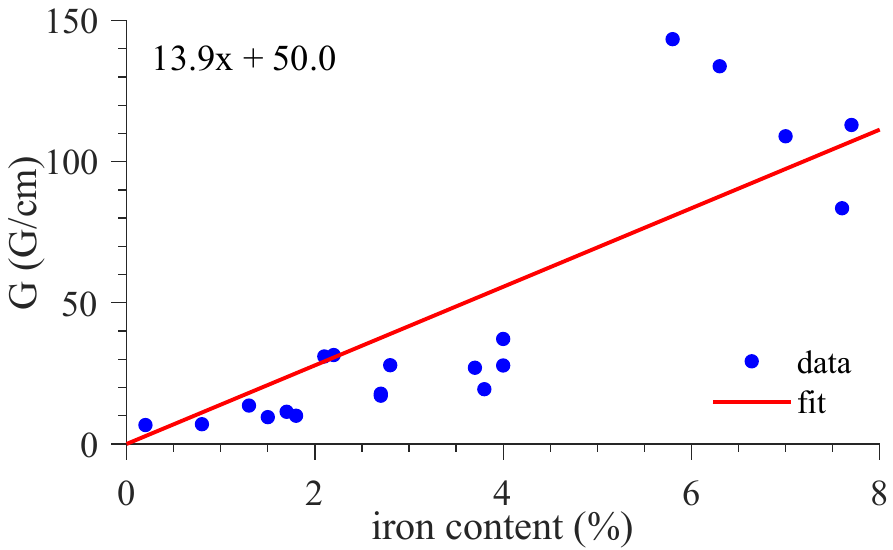}
	\caption{(Color online). Intensity of the internal gradient G as function of the total iron content of the analyzed sandstones.}
\end{figure}

\begin{figure*}
	\centering
	\includegraphics[width=0.9\textwidth]{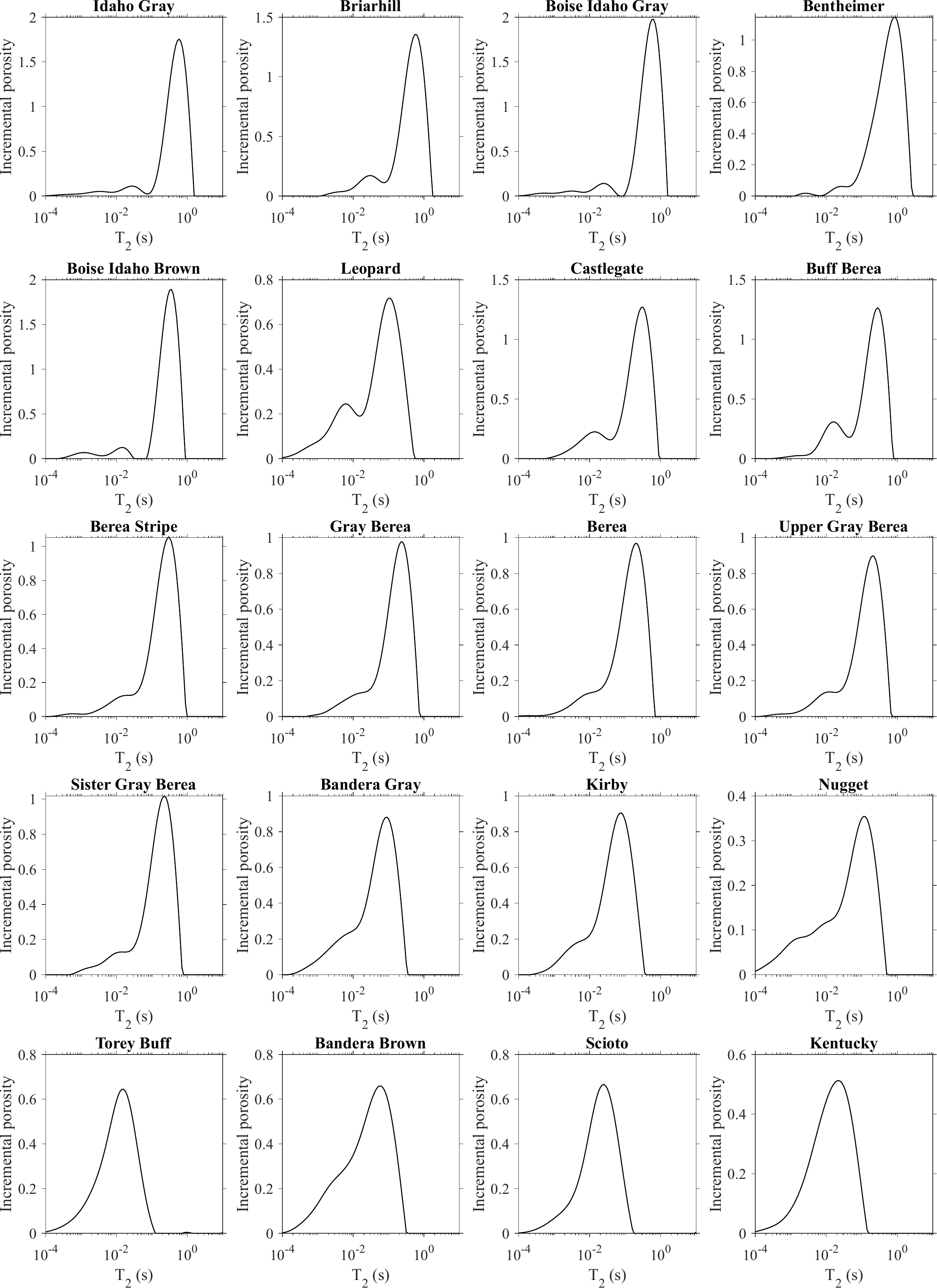}
	\caption{(Color online). \textbf{Distributions of transverse relaxation times.} The T$_2$ distributions for all 20 sandstone samples analyzed, ordered according to their permeability in descending order.}
\end{figure*}

\textbf{Iron and internal gradient intensity.} The amount of iron in a sandstone is directly related to the intensity of the internal gradient inside the pores. These internal gradients are generated due to the differences in the magnetic susceptibility between the matrix and the fluid saturating the porous medium. The intensity of the internal gradients depends not only on the amount iron in the sample, but also on how it is distributed along the pore matrix, with the iron ions near the pore surfaces contributing to the internal gradients. Figure A1 shows the good correlation between the total iron content and the intensity of the internal gradients.

\textbf{Distributions of transverse relaxation times.} The T$_2$ distributions for all 20 sandstone samples we analysed are shown in Figure A2. In order to obtain these distributions, we applied an Inverse Laplace Transform (ILT) on the CPMG decay data, with $\tau =$ 0.1 ms. Each distribution has 512 points, which is necessary to correctly determine the value of T$_{2}^{peak}$ as function of $\tau$, since the ILT procedure introduces some redundancies in the T$_2$ distribution.


%

\end{document}